# Effect of superconductivity by Nb and V substitution in kagome CaPd$_5$

**Dan Li[1], Zhengxuan Wang[1], Chuanguang Zhang[1], Chunlan Ma[2,*], Shijing Gong[3], Chuanxi Zhao[4], Shuaikang Zhang[1], Tianxing Wang[1], Xiao Dong[1], Wuming Liu[5,*] and Yipeng An[1,*]**

[1] School of Physics, Henan Normal University, Xinxiang 453007, China
[2] School of Physics and Technology, Suzhou University of Science and Technology, Suzhou 215009, China
[3] Department of Physics, East China Normal University, Shanghai 200062, China
[4] Department of Physics, Guangdong Provincial Key Laboratory of Nanophotonic Manipulation, Jinan University, Guangzhou 510632, China
[5] Beijing National Laboratory for Condensed Matter Physics, Institute of Physics, Chinese Academy of Sciences, Beijing 100190, China

E-mail: wlxmcl@mail.usts.edu.cn, wliu@iphy.ac.cn, and ypan@htu.edu.cn

## Abstract

Materials featuring kagome lattices have attracted significant research interest due to their unique geometric frustration, which gives rise to rich physical phenomena such as non-trivial topology, spin fluctuations, and superconductivity. In this work, using CaPd$_5$ as the prototype structure, we discover and systematically investigate a new class of kagome superconductors, Ca$M_x$Pd$_{5-x}$ ($M$ = Nb and V) alloys. First-principles calculations confirm that these compounds are non-magnetic metals, among which four are dynamically stable: CaNb$_5$, CaV$_5$, CaNb$_2$Pd$_3$, and CaV$_2$Pd$_3$. CaNb$_5$ is identified as a strong electron-phonon coupling (EPC) superconductor with the highest superconducting transition temperature ($T_c$) of 10.1 K, which can be further increased to 12.8 K under external pressure. In contrast, CaV$_5$, CaNb$_2$Pd$_3$, and CaV$_2$Pd$_3$ exhibit weaker EPC and correspondingly lower $T_c$ values. Furthermore, by applying the method of symmetry indicators, we systematically classify the topological and nodal characteristics of CaNb$_5$, providing valuable insights for determining its superconducting pairing symmetry. Our findings demonstrate that Nb and V substitution in kagome CaPd$_5$ provides an effective route for designing a new type of kagome superconductor with relatively high $T_c$. This study also offers new perspectives on topological superconductivity in kagome systems and establishes a useful guideline for discovering other superconducting materials with unique properties.

## 1. Introduction

Kagome lattices [1, 2] consist of corner-sharing triangular units arranged in a pattern of alternating vertex connectivity. This distinctive geometry gives rise to characteristic electronic structures, including flat bands [3-7], Dirac cones [8-11] at the Brillouin-zone (BZ) corners, and van Hove singularity [12-15], which collectively lead to a range of emergent quantum phenomena. To date, materials hosting kagome lattices have been found to exhibit diverse physical properties, such as quantum spin liquids [16-18], magnetic Weyl fermions [19-21], and chiral anomalies [22-24]. Notably, both theoretical and experimental studies [25, 26] have established kagome systems as a fertile platform for investigating superconductivity and non-trivial topological states.

---

* Author to whom any correspondence should be addressed.

Representative examples include boride-based kagome superconductors such as $MgB_3$ [27-32], intermetallic kagome systems like FeSn [5, 33] and CoSn [34, 35], as well as the recently discovered $AV_3Sb_5$ (*A* = K, Rb, Cs) kagome metals reported by Ortiz *et al*. [36-38]. The continued exploration of novel kagome materials thus remains an endeavor of considerable scientific and technological relevance.

In our recent work, we identified and characterized a series of $AB_5$-type kagome superconductors $MPd_5$ (*M* = Ca, Sr, Ba) [39], all of which exhibit relatively low superconducting transition temperatures. Among them, $CaPd_5$ possesses the highest $T_c$ at 2.64 K, a value that can only be marginally enhanced through external pressure or carrier doping. In this study, we aim to further improve the superconductivity of $CaPd_5$ via elemental substitution, a well-established strategy for synthesizing new materials. Notably, niobium (Nb) [40], a prototype superconductor used in device applications, exhibits the highest $T_c$ (9.2 K) among all elemental superconductors. Motivated by this, we introduce Nb and its same-group element V as substituents in $CaPd_5$, with the dual goal of enhancing $T_c$ and discovering new kagome superconducting systems.

In this work, we first obtain six new kagome materials via elemental substitution in $CaPd_5$ by Nb and V atoms: $CaNb_5$, $CaV_5$, $CaNb_2Pd_3$, $CaV_2Pd_3$, $CaNb_3Pd_2$, and $CaV_3Pd_2$. These six structures are collectively referred to as $CaM_xPd_{5-x}$ (*M* = Nb and V) alloys. Based on first-principles calculations, we systematically evaluate the structural stability of these $CaM_xPd_{5-x}$ systems and investigate their phonon-mediated superconducting properties. We further explore avenues to enhance their superconductivity through the application of external pressure and carrier doping. Finally, we assess their potential for topological superconductivity by analyzing the corresponding pairing symmetries [41, 42].

## 2. Theoretical Methods

This study employs first-principles calculations based on density function theory (DFT), as implemented in the popular open-source QUANTUM ESPRESSO (QE) package [43], to investigate the electronic structures and superconducting properties of $CaM_xPd_{5-x}$ (*M* = Nb and V) alloys. The calculations utilize the Perdew-Burke-Ernzerhof (PBE) generalized gradient approximation [44-47] and SG15 norm-conserving pseudopotentials [48-50]. Structural optimizations are carried out with strict convergence thresholds: total energy changes below $10^{-8}$ Ry and maximum atomic force below $10^{-6}$ Ry Bohr$^{-1}$. The plane-wave kinetic energy cutoffs are set to 80 Ry for wavefunctions and 320 Ry for charge density. A $12^3$ Monkhorst-Pack **k**-mesh is adopted for charge density calculations, while a finer $24^3$ mesh is used for electronic band structure and density of states (DOS) evaluations.

Lattice dynamics are studied within the framework of density-functional perturbation theory [51] using a $6^3$ **k**-mesh and a $3^3$ **q**-mesh in the phonon calculations. The optimized tetrahedron method [52] is applied for Brillouin zone integration and electron-phonon coupling (EPC) calculations. Superconducting properties are evaluated via the density functional theory for superconductors (SCDFT) approach as implemented in the SUPERCONDUCTING-TOOLKIT package [53-58]. Topological characteristics are assessed using the symmetry indicators (SIs) method [59-61], where input files for the subroutine Topological Supercon code [42] are generated from QE self-consistent results post-processed with the QEIRREPS utility [62].

The superconducting transition temperature $T_c$ in this work is numerically determined using the bisection method [54, 63], defined as the temperature at which the superconducting gap Δ drops to zero. The computational procedure starts by solving the gap equation *Eq.* (1) self-consistently. The initial temperature bounds are set as $T_c^{min}$ = 0 K and $T_c^{max}$ is obtained by the Bardeen-Cooper-Schrieffer [64] theory. The gap *Eq.* (1) is then iteratively solved at the midpoint temperature of $T_c^{min}$ and $T_c^{max}$. This process is repeated for 10 cycles under the convergence criterion $\langle|\Delta|\rangle < 10^{-3}\Delta_0$, where $\Delta_0$ denotes the Fermi-surface-averaged gap at 0 K. The final $T_c$ is taken as $(T_c^{min}+T_c^{max})/2$. The **k**-space resolved superconducting gap $\Delta_{n\mathbf{k}}$ is expressed as follows:

$$\Delta_{n\mathbf{k}} = -\frac{1}{2}\sum_{n'\mathbf{k}'} \frac{K_{n\mathbf{k}n'\mathbf{k}'}(\varepsilon_{n\mathbf{k}},\varepsilon_{n'\mathbf{k}'})}{1+Z_{n\mathbf{k}}(\varepsilon_{n\mathbf{k}})} \frac{\Delta_{n'\mathbf{k}'}}{\sqrt{\varepsilon_{n'\mathbf{k}'}^2+\Delta_{n'\mathbf{k}'}^2}}$$
$$\times \tanh\left(\frac{\sqrt{\varepsilon_{n'\mathbf{k}'}^2+\Delta_{n'\mathbf{k}'}^2}}{2T}\right) \qquad (1)$$

The EPC constant $\lambda$ is calculated as the integral of the electron-phonon spectral function $\alpha^2F(\omega)$ according to the relation:

$$\lambda = \sum_{\mathbf{q}\nu}\lambda_{\mathbf{q}\nu} = 2\int_0^\omega \frac{\alpha^2F(\omega)}{\omega}d\omega, \qquad (2)$$

$$\alpha^2F(\omega) = \frac{1}{2\pi N(\varepsilon_F)}\sum_{\mathbf{q}\nu}\delta(\omega-\omega_{\mathbf{q}\nu})\frac{\gamma_{\mathbf{q}\nu}}{\hbar\omega_{\mathbf{q}\nu}}. \qquad (3)$$

## 3. Results and Discussion

### 3.1 Electronic structures of $CaM_xPd_{5-x}$ (*M* = Nb and V)

We begin by constructing and examining the geometrical structures of $CaM_xPd_{5-x}$ alloys, which adopt regular kagome lattice without distortion and share the same symmetry as $CaPd_5$ and $AV_3Sb_5$ (*A* = K, Rb, Cs) systems (i.e., the $D_{6h}$ point group and the *P*6/*mmm* space group) [36-39]. $CaNb_5$ and $CaV_5$ are obtained by replacing all Pd atoms in the kagome layer of $CaPd_5$ with Nb and V atoms, respectively (see figure 1(a)). Similarly, $CaNb_2Pd_3$ and $CaV_2Pd_3$ are formed by substituting Pd atoms in the upper kagome layer with Nb and V atoms, as shown in figures 1(c) and (e). In contrast, $CaNb_3Pd_2$ and $CaV_3Pd_2$ obtained by replacing Pd atoms in the lower kagome layer with Nb or V are dynamically unstable due to the presence of large negative phonon frequencies (see figure S1 in the Supporting Information and the following discussions). Such instability might be ascribed to a charge density wave, or it could potentially be suppressed by applying external pressure, which, for instance, could increase the softened phonon frequency.

**Table 1.** Lattice constants (LC), dynamic stability, the length of triangle in the kagome lattice (Length), the corresponding cumulative electron-phonon coupling strength $\lambda(\omega)$, and superconducting transition temperature $T_c$ of the $CaM_xPd_{5-x}$ ($M$ = Nb and V) alloys.

| System | LC (Å) a/b | LC (Å) c | Stability | Length (Å) | $\lambda(\omega)$ | $T_c$ |
|---|---|---|---|---|---|---|
| $CaNb_5$ | 5.94 | 4.31 | Yes | 2.9707 | 1.17 | 10.1 |
| $CaV_5$ | 5.58 | 3.81 | Yes | 2.7909 | 0.86 | 9.1 |
| $CaNb_2Pd_3$ | 5.53 | 5.53 | Yes | 2.7664 | 0.81 | 5.5 |
| $CaV_2Pd_3$ | 5.46 | 4.14 | Yes | 2.7299 | 0.59 | 3.4 |
| $CaNb_3Pd_2$ | 5.42 | 4.75 | No | / | / | / |
| $CaV_3Pd_2$ | 5.11 | 4.57 | No | / | / | / |

$CaNb_3Pd_2$ and $CaV_3Pd_2$ are expected to become the superconductors under the high-pressure conditions. Therefore, in the subsequent analysis of electronic structures, superconductivity, and topological properties, we focus exclusively on the first four stable configurations: $CaNb_5$, $CaV_5$, $CaNb_2Pd_3$, and $CaV_2Pd_3$. Their lattice constants, dynamical stability, and the lengths of the triangle in the kagome lattice are summarized in table 1. All these compounds exhibit metallic behavior, and their band structures are displayed in figures 1(b), (d), (f), and figure S3(a), respectively. Furthermore, we find that the spin-orbit coupling (SOC) effect is weak in these four structures and does not significantly alter their electronic bands near the Fermi level ($E_F$). Thus, SOC is omitted in the calculations except when investigating topological properties.

The more detailed electronic structures of $CaM_xPd_{5-x}$ are presented in figure 2, including their element- and orbital-projected band structures (fat bands) and density of states (PDOS). For $CaNb_5$, the fat bands and PDOS in figure 2(a) indicate that the $Nb_{4d}$ orbitals dominate the electron states near the $E_F$, representing the sum of all five $4d$ atomic orbitals. Among these, the $d_{xz}$ and $d_{yz}$ orbitals contribute equally, as do the $d_{xy}$ and $d_{x^2-y^2}$ orbitals. The Fermi velocity (FV) projected on the Fermi surfaces for Ca and $Nb_{4d}$ orbitals is shown in figures 2(d) and (e), respectively, with the $Nb_{4d}$ orbitals exhibiting significantly larger contributions. In addition, the FV projections of the $Nb_{4d}$ orbitals for the five bands crossing $E_F$ are provided in figure S2. Note that the electronic structures of $CaV_5$ are similar to those of $CaNb_5$ and are provided in figure S3 in the Supporting Information.

For $CaNb_2Pd_3$ and $CaV_2Pd_3$, both the $Pd_{4d}$ and $Nb_{4d}$ (or $V_{3d}$) orbitals contribute significantly to the states near $E_F$. The fat bands of Nb/V and Pd atoms show slight overlap, suggesting hybridization between their orbitals. As shown in figures 2(b) and (c), the dominant contribution comes from the Nb-$d_{z^2}$ (or V-$d_{z^2}$) orbitals, followed by Nb-$d_{xz}$/V-$d_{xz}$ and Nb-$d_{yz}$/V-$d_{yz}$ orbitals, which contribute equally. The FV projections of the $Nb_{4d}/V_{3d}$ and $Pd_{4d}$ orbitals in $CaNb_2Pd_3/CaV_2Pd_3$ are also provided in figure S4. These results demonstrate that the elements constituting the kagome layers play a critical role in determining the physical properties of the $CaM_xPd_{5-x}$ kagome alloys.

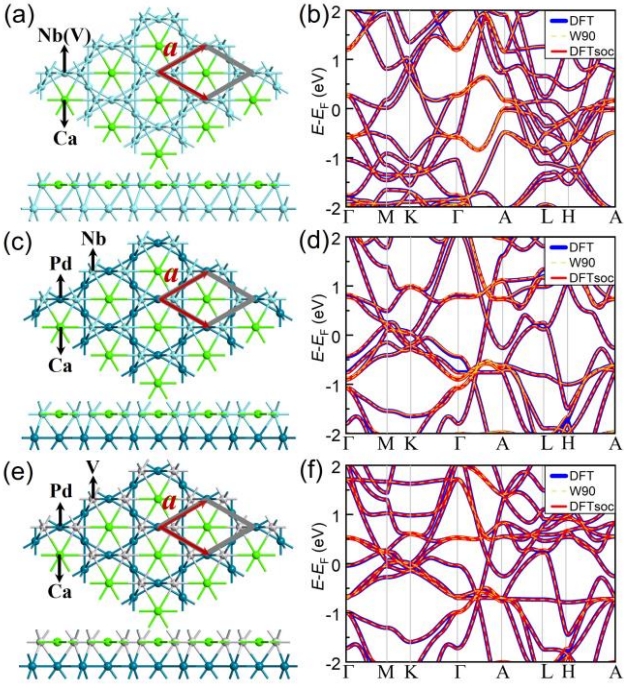

**Figure 1.** Top and side views of (a) $CaNb_5$ ($CaV_5$), (c) $CaNb_2Pd_3$, and (e) $CaV_2Pd_3$. Electron band structures of (b) $CaNb_5$, (d) $CaNb_2Pd_3$, and (f) $CaV_2Pd_3$ obtained by DFT and with spin-orbit coupling (DFTsoc), as well as by tight-binding model *via* wannier90 code [65] (W90). $E_F$ is shifted to zero.

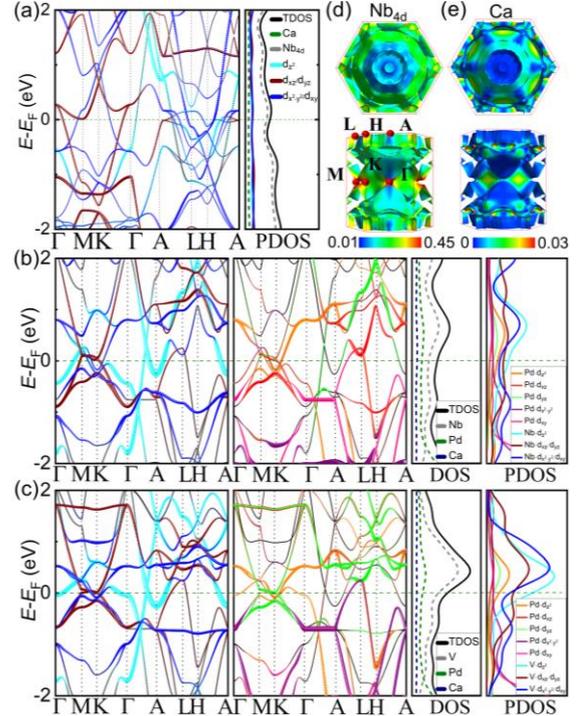

**Figure 2.** Element- and orbital-projected bands and density of states (PDOS) of (a) $CaNb_5$, (b) $CaNb_2Pd_3$, and (c) $CaV_2Pd_3$. Fermi velocity projections on Fermi surfaces of (d) $Nb_{4d}$ orbitals and (e) Ca atoms of $CaNb_5$, obtained by using the FermiSurface package [66]. The six high symmetry positions of first Brillouin-zone of $CaNb_5$ are marked in (d).

### 3.2 Phonon-mediated superconductivity of CaM$_x$Pd$_{5-x}$

We next investigate the phonon properties and phonon-mediated superconductivity in CaM$_x$Pd$_{5-x}$ alloys. Their dynamic stability is assessed through phonon spectrum calculations, as shown in figure 3. The phonon spectra of CaNb$_5$, CaV$_5$, CaNb$_2$Pd$_3$, and CaV$_2$Pd$_3$ exhibit no imaginary frequencies, confirming their dynamical stability. In contrast, CaNb$_3$Pd$_2$ and CaV$_3$Pd$_2$ are dynamically unstable, as evidenced by pronounced phonon softening with relatively large imaginary frequencies. To gain deeper insight into the vibrational properties of CaM$_x$Pd$_{5-x}$ (M = Nb and V), we project the weights of different atomic vibration modes onto the phonon dispersion curves and calculate the phonon density of states (PHDOS). The red and blue bubbles superimposed on the phonon spectra represent the mode-resolved projections of the electron-phonon coupling strength $\lambda_{\mathbf{q}\nu}$ (at wave vector **q** and branch $\nu$) and phonon linewidth $\gamma_{\mathbf{q}\nu}$ (at wave vector **q** and branch $\nu$), respectively. For both CaNb$_5$ and CaV$_5$, the phonon modes across the BZ are dominated by vibrations of Nb or V atoms. In CaNb$_5$, strong EPC strength $\lambda_{\mathbf{q}\nu}$ and phonon linewidths $\gamma_{\mathbf{q}\nu}$ are observed in the out-of-plane acoustic (ZA) branch along the $M$-$\Gamma$ path (see figure 3(a)). Similarly, in CaV$_5$, pronounced $\lambda_{\mathbf{q}\nu}$ and $\gamma_{\mathbf{q}\nu}$ are found in the transverse acoustic (TA) phonon modes (figure 3(b)). These results indicate that low-energy phonon modes, primarily associated with Nb/V atoms and exhibiting a strong EPC, contribute significantly to $\lambda_{\mathbf{q}\nu}$ and $\gamma_{\mathbf{q}\nu}$, and thus play a key role in the phonon-mediated superconductivity of these compounds. For CaNb$_2$Pd$_3$ and CaV$_2$Pd$_3$, phonon modes across the BZ arise from collective vibrations of Pd and Nb/V atoms, as shown in figures 3(c) and (d). Strong EPC $\lambda_{\mathbf{q}\nu}$ is mainly distributed in the ZA modes, while notable phonon linewidths $\gamma_{\mathbf{q}\nu}$ appears around 20 meV, corresponding to high-energy vibrational modes.

The right panels of figures 3(a)-(d) show the Eliashberg spectral function $\alpha^2F(\omega)$ and the cumulative EPC strength $\lambda(\omega)$. For CaNb$_5$, $\alpha^2F(\omega)$ is mainly distributed between 8 and 19 meV, with $\lambda$ reaching 87% of the total EPC constant $\lambda$ at 19 meV. In contrast, CaV$_5$ exhibits dominant $\alpha^2F(\omega)$ contributions in the 11–26 meV range, accounting for 88% of the total $\lambda$ at 26 meV. For CaNb$_2$Pd$_3$ and CaV$_2$Pd$_3$, 89% and 85% of the total $\lambda$ are reached at 20 and 22 meV, respectively. Integration of $\alpha^2F(\omega)$ gives total EPC constants of $\lambda$ = 1.17, 0.86, 0.81, and 0.59 for CaNb$_5$, CaV$_5$, CaNb$_2$Pd$_3$, and CaV$_2$Pd$_3$, respectively, indicating strong EPC in CaNb$_5$ and relatively weak coupling in the other three compounds.

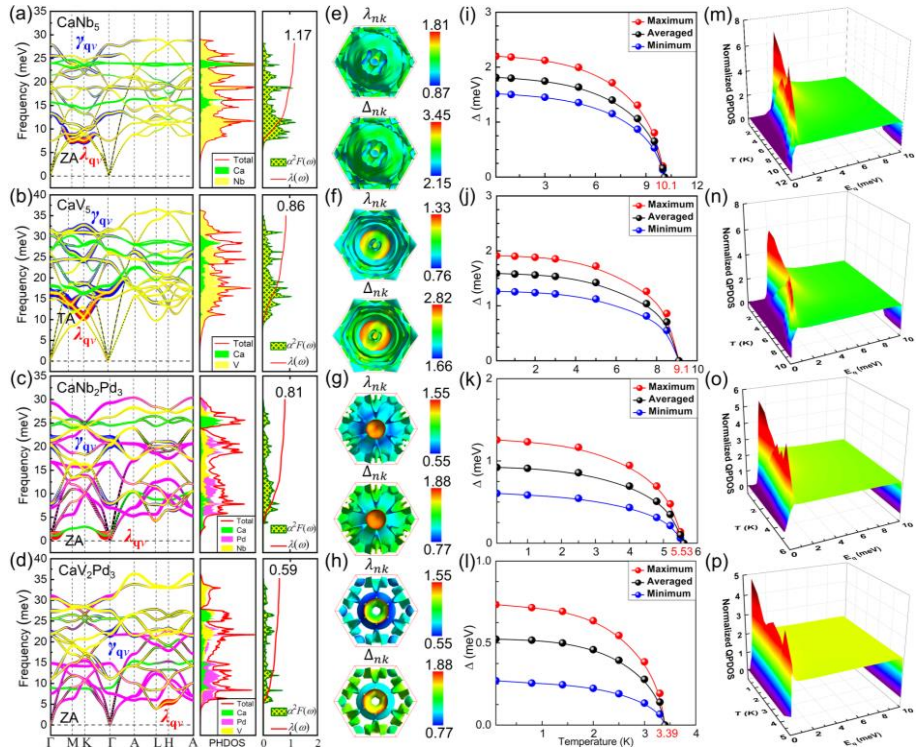

**Figure 3.** Phonon dispersion with electron-phonon coupling $\lambda_{\mathbf{q}\nu}$ (at wavevector **q** and branch $\nu$, red bubbles), phonon linewidth $\gamma_{\mathbf{q}\nu}$ (at wavevector **q** and branch $\nu$, blue bubbles), projections of Ca (green bubbles), Nb/V (yellow bubbles), and Pd (pink bubbles), element-projected phonon density of state (PHDOS), Eliashberg electron-phonon spectral function $\alpha^2F(\omega)$, and cumulative electron-phonon coupling strength $\lambda(\omega)$ for CaM$_x$Pd$_{5-x}$ are presented in panels (a)-(d). Panels (e)-(h) display the Fermi surfaces with top-view projections of electron-phonon coupling strength $\lambda_{n\mathbf{k}}$ and the superconducting gap $\Delta_{n\mathbf{k}}$ (band index $n$ and wave number **k** dependent at 0 K), respectively. The temperature dependence of the superconducting gap $\Delta_{n\mathbf{k}}$ is shown in (i)-(l). The black, red, and blue balls denote the Fermi-surface-averaged, maximum, and minimum values of the gap function at each temperature, with calculations spanning from 0.1 K to $T_c$. (m)-(p) show the normalized quasiparticle density of states (QPDOS) of CaM$_x$Pd$_{5-x}$ at different temperatures.

Figures 3(e)-(h) present the momentum-resolved projections of the EPC strength $\lambda_{n\mathbf{k}}$ and superconducting gap $\Delta_{n\mathbf{k}}$ on the Fermi surfaces, revealing significant anisotropy. In CaNb$_5$, both $\lambda_{n\mathbf{k}}$ and $\Delta_{n\mathbf{k}}$ are also dominated by Nb$_{4d}$ orbitals. More broadly, both $\lambda_{n\mathbf{k}}$ and $\Delta_{n\mathbf{k}}$ in the Ca$M_x$Pd$_{5-x}$ systems exhibit clear anisotropy across the Fermi surfaces. Using the bisection method [53-58], we obtain intrinsic $T_c$ = 10.1, 9.1, 5.3, and 3.4 K for CaNb$_5$, CaV$_5$, CaNb$_2$Pd$_3$, and CaV$_2$Pd$_3$, respectively. As summarized in table 1, $T_c$ is positively correlated with $\lambda$, confirming that these kagome materials are phonon-mediated superconductors.

Moreover, the temperature dependence of the superconducting gap can provide an intuitive means to determine the $T_c$. We further examine the behavior of the superconducting gap function $\Delta_{n\mathbf{k}}$ described by Eq. (1) at different temperatures. Figures 3(i)-(l) display the Fermi-surface-averaged values of the gap function at each temperature, together with the corresponding maximum and minimum values. Notably, all three quantities (the maximum, minimum, and average superconducting gaps) vanish above $T_c$. In addition, each of the normalized quasiparticle density of states (QPDOS) shown in figures 3(m)-(p) exhibits a single peak at 0 K, indicating that the Ca$M_x$Pd$_{5-x}$ ($M$ = Nb and V) systems are single-gap superconductors.

### 3.3 Quantum control of superconductivity of CaNb$_5$

It is well established that external pressure [67-69] and carrier doping [70] are effective means for modulating electronic band structures and superconducting properties. We therefore employ these approaches to enhance the $T_c$ of CaNb$_5$. As shown in figure 4(a), the application of hydrostatic pressure in the range of –10 to 10 GPa induces significant modifications in the band structure of CaNb$_5$. Figures 4(c) and (e) show that lattice constants ($a$ and $c$) decrease monotonically with increasing pressure, while $\lambda$, $\Delta$, and $T_c$ increase monotonically. The $T_c$ of CaNb$_5$ reaches a maximum of 12.8 K at 10 GPa. Notably, beyond 10 GPa, the material enters an unstable phase, as indicated by the appearance of imaginary frequencies in the phonon spectra (see figure S5(a) in the Supporting Information).

To further investigate the pressure effect, we calculate the normalized quasiparticle density of states (QPDOS) at 0.1 K under different pressures (figure 4(g)). The QPDOS exhibits distinct gap features, visualized as a purple region in the figure, which evolve systematically with increasing pressure. Specifically, both the intensity and amplitude of the QPDOS peak are significantly enhanced, accompanied by a shift toward higher energies. These results demonstrate that external pressure can effectively enhance $T_c$ of CaNb$_5$, suggesting a viable route for modulating its superconducting properties.

We further investigate the effects of electron and hole doping by explicitly adjusting the total carrier concentration within a jellium model. Figure 4(b) illustrates the evolution of the electronic band structure under different doping levels, where electron doping shifts the $E_F$ upward, while hole doping shifts it downward. As shown in figure 4(d), the lattice constant L gradually increases with electron doping but decreases monotonically with hole doping. Meanwhile, the $\lambda$, $\Delta$, and $T_c$ exhibit negative correlations with electron doping and positive correlations with hole doping (figure 4(f)). Notably, under a high hole-doping concentration of 0.06 holes per cell, $T_c$ reaches 10.4 K, indicating that hole doping can moderately enhance $T_c$ of CaNb$_5$. Figure 4(h) displays the normalized QPDOS at 0.1 K as a function of carrier concentration. With increasing doping, the QPDOS shows three notable trends: enhanced peak intensity, increased amplitude, and a systematic shift of spectral weight to higher energies. Furthermore, our results reveal that external pressure modulates $T_c$ of CaNb$_5$ more effectively than carrier doping.

### 3.4 Topological superconductivity of Ca$M_x$Pd$_{5-x}$

Finally, we study the pairing symmetry and topological superconductivities in Ca$M_x$Pd$_{5-x}$ using the symmetry indicators method [39, 71], which enables an efficient diagnosis of topological phases through the examination of space group irreducible representations. These systems crystallize in the $D_{6h}$ point group (PG), similar to the hexagonal kagome superconductors $A$V$_3$Sb$_5$ ($A$ = K, Rb, Cs) [36-38]. For CaNb$_5$, our analysis proceeds as follows: building on the self-consistent results from QE code as post-processed by the QEIRREPS utility [62], we subsequently employ the Topological Supercon code [42] to systematically identify the eight time-reversal-symmetric irreducible representations ($B_{1u}$, $B_{1g}$, $B_{2u}$, $B_{2g}$, $A_{2u}$, $A_{2g}$, $A_{1u}$, and $A_{1g}$). The corresponding results are summarized in table 2, while those for other materials are provided in table S1 of the Supplementary Information.

A superconductor is classified as a representation-enforced nodal superconductor (NSC) if it violates specific symmetry constraints known as compatibility relations. By analyzing these violations, the positions and shapes of nodes (e.g., point, line, or surface) can be determined. Conversely, when all compatibility relations are satisfied, the system may be a topological superconductor (TSC) or a topological NSC.

**Table 2.** Topological superconductivity of CaNb$_5$ with $D_{6h}$ point group (PG) symmetry. The second column refers to the irreducible representation [41]. The fourth column is either: the paths where compatibility relations are violated (Case I), or the corresponding entries of symmetry indicators (Case II). Last one means its topological superconductivity. Here, NSC denotes representative-enforced nodal superconductivity, TSC represents symmetry-diagnosable topological superconductor, and TNSC refers to topological NSC.

| PG | Pairing | Case | Nodes | Topology |
|---|---|---|---|---|
| $D_{6h}$ | $B_{1u}$ | I [$P$] | A-L, | NSC |
| | $B_{1g}$ | I [$L$] | A-H, A-L, M-L | NSC |
| | $B_{2u}$ | I [$P$] | $A$-$H$, $K$-$H$ | NSC |
| | $B_{2g}$ | I [$L$] | A-H, A-L, M-L, K-H | NSC |
| | $A_{2u}$ | I [$P$] | $\Gamma$-$A$, $M$-$L$, $K$-$H$ | NSC |
| | $A_{2g}$ | I [$L$] | A-H, A-L, M-L, K-H | NSC |
| | $A_{1u}$ | II | (0, 0, 0, 0, 0, 0, 1, 1, 1, 11, 7) | TSC or TNSC |
| | $A_{1g}$ | IV | … | … |

Based on pairing symmetry analysis, each pairing channel is categorized as: Case I, representation-enforced NSC; Case II, symmetry-diagnosable TSC or topological NSC; Case III, topologically trivial or non-symmetry-diagnosable TSC; or Case IV, silent for trivial pairing (see Ref. [71] for more details). As shown in table 2, for $CaNb_5$, the $B_{1u}$, $B_{1g}$, $B_{2u}$, $B_{2g}$, $A_{2u}$, and $A_{2g}$ pairings are all identified as representation-enforced NSCs (Case I). In contrast, the $A_{1u}$ pairing falls into Case II, suggesting it could be a symmetry-diagnosable TSC or topological NSC (TNSC). Notably, the $A_{1g}$ pairing corresponds to Case IV, for which the current method lacks predictive power.

The predicted results indicate that the superconducting states are predominantly representation-enforced NSCs, with the exception of the $A_{1u}$ or $A_{1g}$ pairing symmetry. Regarding the nodal structures, odd-parity pairings host point nodes, while even-parity pairings exhibit line nodes. This behavior is consistent with that observed in the kagome superconductor $CsV_3Sb_5$ [71]. Moreover, for the specific case of $A_{1u}$ pairing, its topological properties can be directly diagnosed using the SIs method [39]. The results show that the material corresponds to the entry $(0, 0, 0, 0, 0, 1, 1, 1, 11, 7) \in (Z_1)^4 \times (Z_2)^4 \times (Z_{12})^4 \times (Z_{24})^4$ of SIs, indicating that the mirror Chern number on the $k_z = \pi$ plane is 1 mod 2, while on the $k_z = 0$ plane it is 7 mod 12. This approach allows us to characterize the topological superconductivity in $CaM_xPd_{5-x}$ kagome materials and identify their likely pairing symmetries. The pairing symmetry predictions made by this method can be directly compared with experimental results, thereby facilitating a more rapid confirmation of the superconducting pairing symmetry.

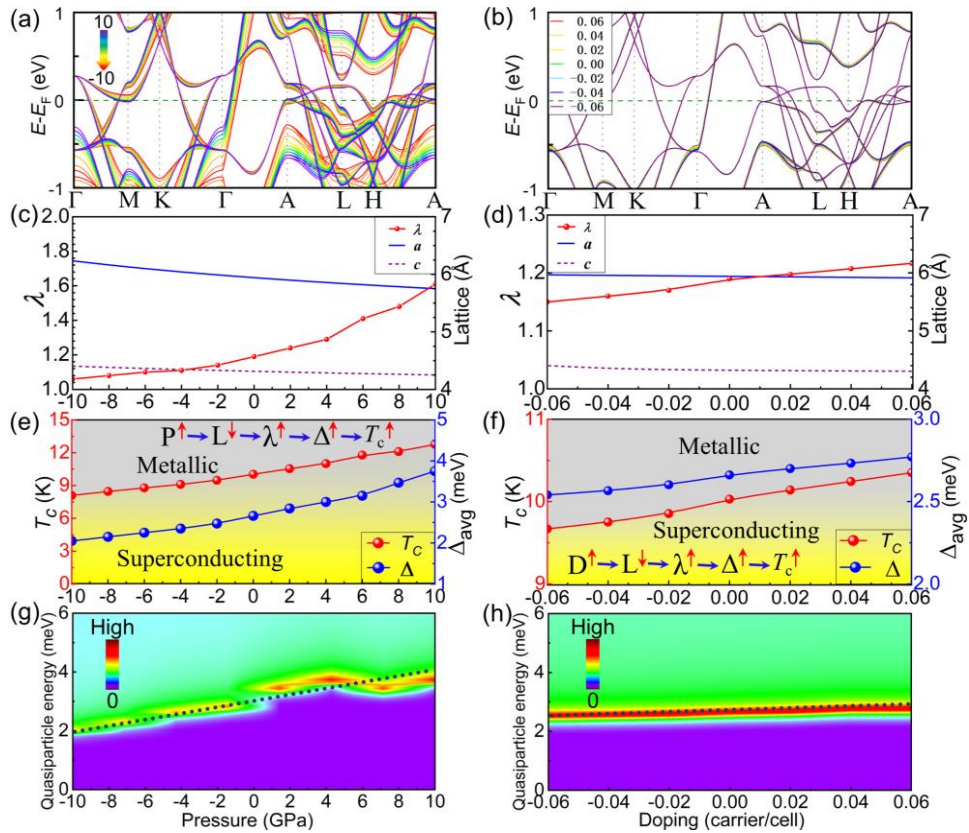

**Figure 4.** Electron band of $CaNb_5$ under hydrostatic pressures (a) and varying carrier concentrations (b). The electron-phonon coupling $\lambda$ and lattice parameters (L = $a$ and $c$) under various pressures (c) and carrier (hole and electron) doping concentrations (d). $T_c$ and superconducting gap $\Delta$ as functions of pressures (e) and carrier doping concentrations (f). The phase diagram distinguishes metallic (gray shaded) and superconducting (yellow shaded) regions. Normalized quasiparticle density of states at 0.1 K as a function of pressure (g) and carrier doping concentrations (h). The black dotted lines track the shift of the single-gap peak.

## 4. Conclusion

In conclusion, based on first-principles calculations, we predict a new family of kagome superconductors, $CaM_xPd_{5-x}$ ($M$ = Nb and V), obtained by elemental substitution of Nb or V in the prototype compound $CaPd_5$. Phonon spectrum calculations confirm the dynamical stability of four structures: $CaNb_5$, $CaV_5$, $CaNb_2Pd_3$, and $CaV_2Pd_3$. Using density functional theory for superconductors, we find that $CaNb_5$ is a strong electron-phonon coupling ($\lambda > 1$) superconductor with the highest superconducting transition temperature $T_c$ among the series, while $CaNb_5$, $CaNb_2Pd_3$, and $CaV_2Pd_3$ exhibit weak electron-phonon coupling ($\lambda < 1$), with respective $\lambda$ of 1.17, 0.86, 0.81, and 0.59. The corresponding superconducting transition temperatures $T_c$ are 10.1, 9.1, 5.5, and 3.4 K, respectively.

We further explore the effects of hydrostatic pressure and carrier doping on enhancing the superconductivity of CaNb$_5$. Under applied pressure, its $T_c$ rises to 12.8 K, and hole doping at a concentration of 0.06 holes per unit cell increases $T_c$ to 10.4 K. Compared to the recently predicted $M$Pd$_5$ ($M$ is a group-IIA metal element) family kagome superconductors, the Ca$M_x$Pd$_{5-x}$ systems offer a dual advantage: a higher $T_c$ and lower synthesis costs due to a lowered reliance on palladium. Topological superconductivity in these materials is also investigated *via* symmetry indicators of pairing symmetry and mirror Chern numbers. Our findings suggest that Ca$M_x$Pd$_{5-x}$ kagome materials are promising platforms for exploring topological superconductivity and provide valuable insights for the discovery of other kagome superconductors with distinctive physical properties. We also hope that our theoretical findings may motivate further experimental investigations on this family of kagome AB$_5$-type superconductors.

## Data availability statement

All data that support the findings of this study are included within the article (and any supplementary files).

## Acknowledgements

This work was supported by the Natural Science Foundation of Henan (Grant No. 242300421214), the National Natural Science Foundation of China (Grant No. 12274117), the Program for Innovative Research Team (in Science and Technology) in the University of Henan Province (Grant No. 24IRTSTHN025), the Open Fund of Guangdong Provincial Key Laboratory of Nanophotonic Manipulation (Grant No. 202502), the Guangdong S&T Program (Grant No. 2023B1212010008), and the HPCC of HNU. We also acknowledge financial support from the National Natural Science Foundation of China (Grants No. 62274066, No. 62275074, and No. 12504129). The authors acknowledge the assistance of DeepSeek tool in language polishing during the manuscript preparation.

## ORCID iDs

Y P An https://orcid.org/0000-0001-5477-4659
W M Liu https://orcid.org/0000-0002-1179-2061
C L Ma https://orcid.org/0000-0002-9944-4255

## References

[1] Syôzi I 1951 Statistics of kagomé lattice *Prog. Theor. Phys.* **6** 306-308
[2] Mekata M 2003 Kagome: The story of the basketweave lattice *Phys. Today* **56** 12-13
[3] Lin Z *et al* 2018 Flatbands and emergent ferromagnetic ordering in Fe$_3$Sn$_2$ kagome lattices *Phys. Rev. Lett.* **121** 096401
[4] Kang M *et al* 2020 Topological flat bands in frustrated kagome lattice CoSn *Nat. Commun.* **11** 4004
[5] Kang M *et al* 2020 Dirac fermions and flat bands in the ideal kagome metal FeSn *Nat. Mater.* **19** 163-169
[6] Meier W R, Du M H, Okamoto S, Mohanta N, May A F, McGuire M A, Bridges C A, Samolyuk G D and Sales B C 2020 Flat bands in the CoSn-type compounds *Phys. Rev. B* **102** 075148
[7] Sethi G, Zhou Y, Zhu L, Yang L and Liu F 2021 Flat-band-enabled triplet excitonic insulator in a diatomic kagome lattice *Phys. Rev. Lett.* **126** 196403
[8] Liu Z *et al* 2020 Orbital-selective dirac fermions and extremely flat bands in frustrated kagome-lattice metal CoSn *Nat. Commun.* **11** 4002
[9] Mizoguchi T, Kuno Y and Hatsugai Y 2021 Flat band, spin-1 dirac cone, and hofstadter diagram in the fermionic square kagome model *Phys. Rev. B* **104** 035161
[10] Hu Y, Wu X, Yang Y, Gao S, Plumb N C, Schnyder A P, Xie W, Ma J and Shi M 2022 Tunable topological dirac surface states and van hove singularities in kagome metal GdV$_6$Sn$_6$ *Sci. Adv.* **8** eadd2024
[11] Kumar Pradhan S, Pradhan S, Mal P, Rambabu P, Lakhani A, Das B, Lingam Chittari B, Turpu G R and Das P 2024 Endless dirac nodal lines and high mobility in kagome semimetal Ni$_3$In$_2$Se$_2$: a theoretical and experimental study *J. Phys.:Condens. Matter* **36** 445601
[12] Wang W-S, Li Z-Z, Xiang Y-Y and Wang Q-H 2013 Competing electronic orders on kagome lattices at van hove filling *Phys. Rev. B* **87** 115135
[13] Gao Q, Yan Q, Hu Z and Chen L 2024 Bilayer kagome borophene with multiple van hove singularities *Adv. Sci.* **11** 2305059
[14] Luo Y *et al* 2023 A unique van hove singularity in kagome superconductor CsV$_{3-x}$Ta$_x$Sb$_5$ with enhanced superconductivity *Nat. Commun.* **14** 3819
[15] Liu B *et al* 2023 Tunable van hove singularity without structural instability in kagome metal CsTi$_3$Bi$_5$ *Phys. Rev. Lett.* **131** 026701
[16] Kelly Z A, Gallagher M J and McQueen T M 2016 Electron doping a kagome spin liquid *Phy. Rev. X* **6** 041007
[17] Schaffer R, Huh Y, Hwang K and Kim Y B 2017 Quantum spin liquid in a breathing kagome lattice *Phys. Rev. B* **95** 054410
[18] Liu Z-X, Tu H-H, Wu Y-H, He R-Q, Liu X-J, Zhou Y and Ng T-K 2018 Non-abelian S=1 chiral spin liquid on the kagome lattice *Phys. Rev. B* **97** 195158
[19] Ren Z *et al* 2022 Plethora of tunable Weyl fermions in kagome magnet Fe$_3$Sn$_2$ thin films *npj Comput. Mater.* **7** 109
[20] Wu M-X, Wei Y-H, Ma D-S, Wang P, Gao N, Wu S-Y and Kuang M-Q 2023 In-plane canted ferromagnetism, intrinsic weyl fermions, and large anomalous hall effect in the kagome semimetal Rh$_3$Sn$_2$S$_2$ *Phys. Rev. B* **108** 174430
[21] Belopolski I *et al* 2021 Signatures of weyl fermion

annihilation in a correlated kagome Magnet *Phys. Rev. Lett.* **127** 256403
[22] Zhang Z-D, Lu M-H and Chen Y-F 2024 Observation of free-boundary-induced chiral anomaly bulk states in elastic twisted kagome metamaterials *Phys. Rev. Lett.* **132** 086302
[23] van Heumen E 2021 Kagome lattices with chiral charge density *Nat. Mater.* **20** 1308-1309
[24] Yu S L and Li J X 2012 Chiral superconducting phase and chiral spin-density-wave phase in a hubbard model on the kagome lattice *Phys. Rev. B* **85** 144402
[25] Mandal M, Kataria A, Meena P K, Kushwaha R K, Singh D, Biswas P K, Stewart R, Hillier A D and Singh R P 2025 Time-reversal symmetry breaking in a Re-based kagome lattice superconductor *Phys. Rev. B* **111** 054511
[26] Meena P K, Mandal M, Manna P, Srivastava S, Sharma S, Mishra P and Singh R P 2024 Superconductivity in breathing kagome-structured C14 Laves phase $XOs_2$ (X = Zr, Hf) *Supercond. Sci. Technol.* **37** 075004
[27] Yang L, Li Y P, Liu H D, Jiao N, Ni M Y, Lu H Y, Zhang P and Ting C S 2023 Theoretical prediction of superconductivity in boron kagome monolayer: $MB_3$ (M = Be, Ca, Sr) and the hydrogenated $CaB_3$ *Chin. Phys. Lett.* **40** 017402
[28] Xie S Y, Li X B, Tian W Q, Chen N K, Wang Y L, Zhang S B and Sun H B 2015 A novel two-dimensional $MgB_6$ crystal: metal-layer stabilized boron kagome lattice *Phys. Chem. Chem. Phys.* **17** 1093-1098
[29] Zheng Q *et al* 2016 Ternary borides $Nb_7Fe_3B_8$ and $Ta_7Fe_3B_8$ with kagome-type iron framework *Dalton Trans.* **45** 9590-9600
[30] Bo T, Liu P F, Yan L and Wang B T 2020 Electron-phonon coupling superconductivity in two-dimensional orthorhombic $MB_6$ (M = Mg, Ca, Ti, Y) and hexagonal $MB_6$ (M = Mg, Ca, Sc, Ti) *Phys. Rev. Mater.* **4** 114802
[31] Qu Z Y, Han F J J, Yu T, Xu M L, Li Y W and Yang G C 2020 Boron kagome-layer induced intrinsic superconductivity in a $MnB_3$ monolayer with a high critical temperature *Phys. Rev. B* **102** 075431
[32] An Y *et al* 2023 Topological and nodal superconductor kagome magnesium triboride *Phys. Rev. Mater.* **7** 014205
[33] Meier W R, Yan J, McGuire M A, Wang X P, Christianson A D and Sales B C 2019 Reorientation of antiferromagnetism in cobalt doped FeSn *Phys. Rev. B* **100** 184421
[34] Sales B C *et al* 2021 Tuning the flat bands of the kagome metal CoSn with Fe, In, or Ni doping *Phys. Rev. Mater.* **5** 044202
[35] Wan S, Lu H and Huang L 2022 Temperature dependence of correlated electronic states in the archetypal kagome metal CoSn *Phys. Rev. B* **105** 155131
[36] Ortiz B R *et al* 2019 New kagome prototype materials: discovery of $KV_3Sb_5$, $RbV_3Sb_5$, and $CsV_3Sb_5$ *Phys. Rev. Mater.* **3** 094407
[37] Ortiz B R, Sarte P M, Kenney E M, Graf M J, Teicher S M L, Seshadri R and Wilson S D 2021 Superconductivity in the $Z_2$ kagome metal $KV_3Sb_5$ *Phys. Rev. Mater.* **5** 034801
[38] Ortiz B R *et al* 2020 $CsV_3Sb_5$: a $Z_2$ topological kagome metal with a superconducting ground state *Phys. Rev. Lett.* **125** 247002
[39] Li D *et al* 2025 $MPd_5$ kagome superconductors studied by density functional calculations *Phys. Rev. B* **111** 144511
[40] Hirshfeld A T, Leupold H A and Boorse H A 1962 Superconducting and normal specific heats of niobium *Phys. Rev.* **127** 1501-1507
[41] Ono S and Shiozaki K 2022 Symmetry-based approach to superconducting nodes: unification of compatibility conditions and gapless point classifications *Phys. Rev. X* **12** 011021
[42] Ono S, Watanabe H, Tang F and Wan X G Topological Supercon (2021) http://toposupercon.t.u-tokyo.ac.jp/tms/
[43] Giannozzi P *et al* 2017 Advanced capabilities for materials modelling with Quantum ESPRESSO *J. Phys.:Condens. Matter* **29** 465901
[44] Perdew J P, Chevary J A, Vosko S H, Jackson K A, Pederson M R, Singh D J and Fiolhais C 1992 Atoms, molecules, solids, and surfaces: applications of the generalized gradient approximation for exchange and correlation *Phys. Rev. B* **46** 6671-6687
[45] Perdew J P, Burke K and Ernzerhof M 1996 Generalized gradient approximation made simple *Phys. Rev. Lett.* **77** 3865-3868
[46] Thonhauser T, Zuluaga S, Arter C A, Berland K, Schröder E and Hyldgaard P 2015 Spin signature of nonlocal correlation binding in metal-organic frameworks *Phys. Rev. Lett.* **115** 136402
[47] Shan W, Shi A, Xin Z, Zhang X, Wang B, Li Y and Niu X 2025 Suppressing the vdw gap-induced tunneling barrier by constructing interfacial covalent bonds in 2D metal–semiconductor contacts *Adv. Funct. Mater.* **35** 2412773
[48] Hamann D R 2013 Optimized norm-conserving vanderbilt pseudopotentials *Phys. Rev. B* **88** 085117
[49] Schlipf M and Gygi F 2015 Optimization algorithm for the generation of ONCV pseudopotentials *Comput. Phys. Commun.* **196** 36-44
[50] Prandini G, Marrazzo A, Castelli I E, Mounet N and Marzari N 2018 Precision and efficiency in solid-state pseudopotential calculations *npj Comput. Mater.* **4** 72
[51] Baroni S, de Gironcoli S, Dal Corso A and Giannozzi P 2001 Phonons and related crystal properties from density-functional perturbation theory *Rev. Mod. Phys.* **73** 515-562
[52] Kawamura M, Gohda Y and Tsuneyuki S 2014 Improved tetrahedron method for the brillouin-zone integration applicable to response functions *Phys. Rev. B* **89** 094515
[53] Sanna A, Pellegrini C and Gross E K U 2020 Combining eliashberg theory with density functional theory for the accurate prediction of superconducting transition temperatures and gap functions *Phys. Rev. Lett.* **125** 057001

[54] Kawamura M, Hizume Y and Ozaki T 2020 Benchmark of density functional theory for superconductors in elemental materials *Phys. Rev. B* **101** 134511

[55] Flores-Livas J A and Sanna A 2015 Superconductivity in intercalated group-IV honeycomb structures *Phys. Rev. B* **91** 054508

[56] Lüders M, Marques M A L, Lathiotakis N N, Floris A, Profeta G, Fast L, Continenza A, Massidda S and Gross E K U 2005 Ab initio theory of superconductivity. I. Density functional formalism and approximate functionals *Phys. Rev. B* **72** 024545

[57] Marques M A L, Lüders M, Lathiotakis N N, Profeta G, Floris A, Fast L, Continenza A, Gross E K U and Massidda S 2005 Ab initio theory of superconductivity. II. Application to elemental metals *Phys. Rev. B* **72** 024546

[58] Oliveira L N, Gross E K U and Kohn W 1988 Density-functional theory for superconductors *Phys. Rev. Lett.* **60** 2430-2433

[59] Slager R-J, Mesaros A, Juričić V and Zaanen J 2013 The space group classification of topological band-insulators *Nat. Phys.* **9** 98-102

[60] Po H C, Vishwanath A and Watanabe H 2017 Symmetry-based indicators of band topology in the 230 space groups *Nat. Commun.* **8** 50

[61] Kruthoff J, de Boer J, van Wezel J, Kane C L and Slager R-J 2017 Topological classification of crystalline insulators through band structure combinatorics *Phys. Rev. X* **7** 041069

[62] Matsugatani A, Ono S, Nomura Y and Watanabe H 2021 qeirreps: An open-source program for Quantum ESPRESSO to compute irreducible representations of bloch wavefunctions *Comput. Phys. Commun.* **264** 107948

[63] An Y *et al* 2021 Superconductivity and topological properties of $MgB_2$-type diborides from first principles *Phys. Rev. B* **104** 134510

[64] Bardeen J C, L. N. Schrieffer, J. R. 1957 Theory of superconductivity *Phys. Rev.* **108** 1175-1204

[65] Mostofi A A, Yates J R, Lee Y-S, Souza I, Vanderbilt D and Marzari N 2008 wannier90: A tool for obtaining maximally-localised wannier functions *Comput. Phys. Commun.* **178** 685-699

[66] Kawamura M 2019 FermiSurfer: Fermi-surface viewer providing multiple representation schemes *Comput. Phys. Commun.* **239** 197-203

[67] Guo J *et al* 2017 Robust zero resistance in a superconducting high-entropy alloy at pressures up to 190 GPa *Proc. Natl. Acad. Sci. U.S.A.* **114** 13144-13147

[68] Majumdar A *et al* 2020 Interplay of charge density wave and multiband superconductivity in layered quasi-two-dimensional materials: the case of $2H\text{-}NbS_2$ and $2H\text{-}NbSe_2$ *Phys. Rev. Mater.* **4** 084005

[69] Jiang K-Y, Cao Y-H, Yang Q-G, Lu H-Y and Wang Q-H 2025 Theory of pressure dependence of superconductivity in bilayer nickelate $La_3Ni_2O_7$ *Phys. Rev. Lett.* **134** 076001

[70] An Y *et al* 2023 Higher-order topological and nodal superconducting transition-metal sulfides MS (M = Nb and Ta) *Phys. Rev. B* **108** 054519

[71] Tang F *et al* 2022 High-throughput investigations of topological and nodal superconductors *Phys. Rev. Lett.* **129** 027001